# GHz optomechanical resonators with high mechanical *Q* factor in air

Xiankai Sun, King Y. Fong, Chi Xiong, Wolfram H. P. Pernice, and Hong X. Tang*

*Department of Electrical Engineering, Yale University,15 Prospect St., New Haven, CT 06511,USA*
*\*hong.tang@yale.edu*

**Abstract:** We demonstrate wheel-shaped silicon optomechanical resonators for resonant operation in ambient air. The high finesse of optical whispering gallery modes (loaded optical *Q* factor above 500,000) allows for efficient transduction of the wheel resonator's mechanical radial contour modes of frequency up to 1.35 GHz with high mechanical *Q* factor around 4,000 in air.

## 1. Introduction

Micro- and nano-optomechanics has recently attracted great interest because of its promise for both practical applications, ranging from mechanical sensing [1–3] to signal processing [4,5], and fundamental research of mesoscale quantum mechanics [6,7]. Obtaining high resonance frequencies in optomechanical devices is desirable for developing high-speed sensing systems, routing signals of different frequencies in optical channels, and also for facilitating access to the quantum regime. At the same time, achieving high mechanical quality ($Q$) factors is equally important, because the readout sensitivity and the coherence time of the mechanical vibration directly scale with the $Q$ factor.

It is well known that the angular mechanical frequency can be expressed as $W_m = (k/m_{eff})^{1/2}$, where $k$ denotes the modal spring constant and $m_{eff}$ is the effective modal mass. Therefore one way to obtain high frequency optomechanical resonators is minimizing $m_{eff}$. This strategy has recently been implemented in silicon (Si) nanobeams [8], optomechanical crystals [9], and GaAs disks [10], where the resulting $m_{eff}$ is reduced to a few picograms or even less. The other route to high mechanical frequency consists in exploring vibrating modes that possess very high stiffness. This latter approach is particularly attractive because it often simultaneously results in high mechanical $Q$ factors. For example, the radial contour modes in micromechanical "hollow-disk" ring resonators are reported to have mechanical $Q$ factors above 10,000 [11,12]. The nonintrusive suspension in these spoke-supported resonators provides significant loss reduction compared with devices where the resonators are directly anchored to the underlying substrate.

Optomechanical resonators with supporting-spoke features were reported previously. They operated either in the flexural mode actuated by the gradient force between the double rings [13,14] or only in the lowest-order radial contour mode (i.e., the radial breathing mode) [12]. As a result, their frequencies were in the range of a few to tens of MHz. In this work, we design, fabricate, and characterize wheel-shaped optomechanical resonators that operate at GHz frequency with high mechanical $Q$ factor in ambient air. Our devices are fabricated on a CMOS-compatible all-integrated Si photonics platform. Compared with fiber taper coupling schemes employed previously [15], our integrated circuit approach provides an efficient framework for characterizing such optomechanical devices because it relaxes stringent fiber alignment requirements and removes the vibrational noise introduced by the fiber. Furthermore, critical coupling of the input waveguide to the cavity's optical modes can be reliably achieved through lithographic alignment. With a high loaded optical $Q$ factor of over 500,000, the wheel's mechanical radial contour modes up to the 4th order (frequency 1.35 GHz) are measured, with high mechanical $Q$ factors around 4,000 at room temperature and atmospheric pressure.

## 2. Mechanical design

We adopt the mechanical structures presented in [11] for building our optomechanical resonators because of their potential in achieving high mechanical $Q$ factors in high-frequency radial contour modes. Studied first by Stephenson, the frequencies of the radial contour modes are determined by the following characteristic equation [16]:

$$F(r_i, r_o, f) = [(\sigma-1)J_1(hr_i) + hr_i J_0(hr_i)] \cdot [(\sigma-1)Y_1(hr_o) + hr_o Y_0(hr_o)] \\ - [(\sigma-1)Y_1(hr_i) + hr_i Y_0(hr_i)] \cdot [(\sigma-1)J_1(hr_o) + hr_o J_0(hr_o)] = 0, \quad (1)$$

where $h = 2\pi f \sqrt{\rho(1-\sigma^2)/E}$, $r_i$ and $r_o$ are the inner and outer radius of the ring and $f$ is the modal vibration frequency. r, s, and $E$ are the density, the Poisson ratio, and the Young's modulus of the material, respectively. $J_n$ and $Y_n$ are Bessel functions of the first and second kinds, respectively. For single-crystal Si, we use the following parameters for theoretical design and numerical simulation: r = 2329 kg/m$^3$, s = 0.28, $E$ = 170 GPa. Figure 1(a) illustrates the solutions of Eq. (1) for an inner radius $r_i$ of 21.22 mm (a number chosen based on the guidelines described below), each line corresponding to a radial contour mode. The displacement profiles obtained from finite-element method (FEM) simulations for each mode are shown in Fig. 1(b). The 1st mode, whose frequency has little dependence on $r_o$ with a fixed $r_i$, is the radial breathing mode due to the collective motion of the ring. All the higher-order modes, with their frequency strongly dependent on the ring width ($r_o$ - $r_i$), are radial pinch modes due to the relative motion inside the ring.

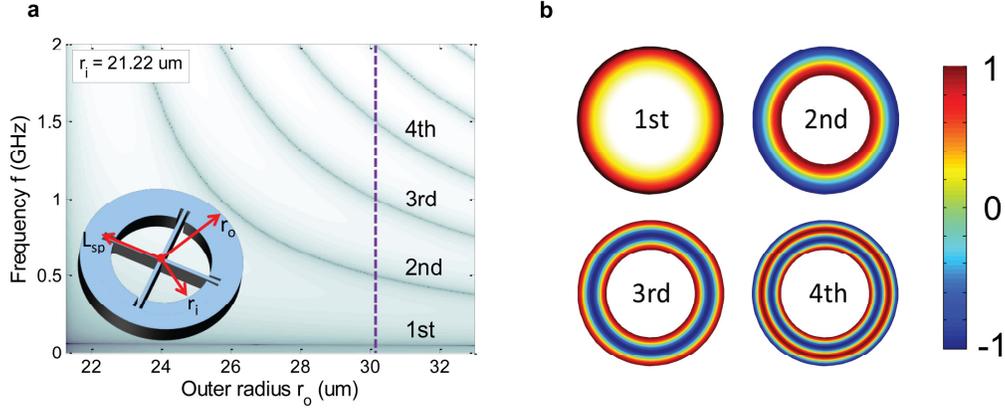

Fig. 1. (a) Contour map showing the solutions of the characteristic equation [Eq. (1)] that determines the frequency of the radial contour modes. The inner radius $r_i$ of the ring is fixed at 21.22 mm. Inset: schematic of the wheel resonator. The vertical dashed line marks the choice of the outer radius $r_o$ = 30.14 mm in our design. (b) Normalized displacement profiles of the first four radial contour modes.

Table 1. Properties of the first four mechanical radial contour modes, theoretical (superscript $t$), simulated (superscript $s$), and experimental (superscript $e$)

| Mode | $W_m^t/2p$ (MHz) | $W_m^s/2p$ (MHz) | $W_m^e/2p$ (MHz) | $m_{eff}^s$ (pg) | $k^s$ ($10^6$ N/m) | $Q_m^e$ |
|---|---|---|---|---|---|---|
| 1st | 53.5 | 48.9 | 46.0 | 324.1 | 0.0306 | 1500 (air), 3000 (vacuum) |
| 2nd | 503 | 466.6 | 465 | 241.4 | 2.075 | 3900 (air), 10000 (vacuum) |
| 3rd | 1000 | 923.8 | 921 | 115.7 | 3.898 | 4100 (air) |
| 4th | 1498 | 1341 | 1347 | 78.4 | 5.566 | 3900 (air) |

As seen in Eq. (1), for a given material, the frequencies $f$ of the radial contour modes are exclusively determined by $r_i$ and $r_o$, and do not depend on the thickness of the geometry in first order approximation. This feature provides great facility in design and fabrication because devices of different resonance frequencies can easily be integrated on the chip with a single run. In order to maximize the mechanical $Q$ factor of a specific mode, we follow two design guidelines when determining the wheel's geometric parameters [11], which are $r_i$ and $r_o$, the inner and outer radius of the ring, and $L_{sp}$, the spoke length [see inset of Fig. 1(a)].

1) To minimize the anchor loss to the substrate, the spoke length $L_{sp}$, defined as the distance from the wheel center to the ring attaching point, should be an odd multiple of a quarter wavelength of the spoke's longitudinal mode:

$$L_{sp} = (2n-1)\frac{\lambda_a}{4} = \frac{2n-1}{4f}\sqrt{\frac{E}{\rho}}, \quad n = 1, 2, 3 \ldots \quad (2)$$

2) To minimize the energy loss from the vibrating ring to the spokes, the attaching point should be located at a nodal point of the targeted radial contour mode. This condition is used to determine a unique set of $(r_i, r_o)$ that both satisfies the frequency condition and has the first maximum of $F(r_i, r_o)$ located at this attaching point.

Following the principles outlined above we designed the wheel structure such that the 3rd mode has a frequency around 1 GHz, which is the target mode for optimization. The spoke length was chosen to be $L_{sp} = 11l_a/4 = 23.50$ mm, and the inner and outer radius are $r_i = 21.22$ mm and $r_o = 30.14$ mm. With these design parameters, the theoretical and simulated modal frequency $W_m/2p$, together with the simulated modal effective mass $m_{eff}$ and modal spring constant $k$, is shown in Table 1 for the first four radial contour modes.

## 3. Device fabrication, measurements, and discussion

We fabricated the optimized device geometry from standard silicon-on-insulator substrates, with a 220-nm silicon layer on 3-mm buried oxide. The entire structure was patterned with high-resolution electron-beam lithography and etched with plasma dry etching based on chlorine chemistry. The wheel resonator was then released from the substrate by photolithography and subsequent wet etching in a buffered oxide etchant, followed by critical point drying. As shown in Fig. 2(a), a typical device includes a pair of grating couplers for vertically coupling light onto and out of the chip, a wheel-shaped optomechanical resonator, and a pulley-shaped wrap-around waveguide for in-plane coupling to the wheel resonator.

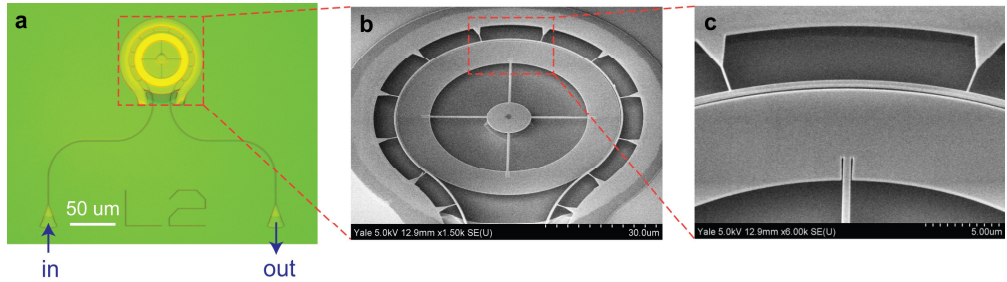

Fig. 2. (a) Top-view optical microscope image showing the entire device, including the grating couplers, the wrap-around coupling waveguide, and the wheel resonator. (b) 45°-tilted-view scanning electron microscope (SEM) image of the wheel resonator. (c) SEM close-up view of the notched attaching point and the wrap-around coupling waveguide.

Because during the release process any features in proximity of the wheel resonator would also inevitably be released from the substrate, the wrap-around waveguide was designed to connect to a large surrounding pad through a series of triangle-shaped anchors to avoid falling off. The pulley-shaped wrap-around waveguide was specially designed to enhance the coupling strength between the waveguide and the wheel resonator. Compared with the traditional point-contact coupling scheme [17], this design allows for a larger waveguide–resonator gap, making the lithography step less challenging [18], and also facilitates excitation and sensing of the mechanical radial contour modes. The actual coupling strength is determined by the geometries of the waveguide and the wheel, and the gap between them. The critical coupling condition was found for a gap of 120 nm between the coupling waveguide (width 500 nm) and the wheel resonator.

Using a tunable diode laser and a low-noise photodetector we characterized the fabricated devices optically. Figure 3(a) shows the transmission spectrum of a device, displaying a good extinction ratio of 15–20 dB. Note that the background fringes result from back-reflection between the two grating couplers, which also contribute a coupling loss of around 8 dB, each. The measured free spectral range of the wheel resonator is 3.39 nm at a wavelength of 1525 nm. Under the assumption that the difference between the whispering gallery modes' circulating radius and the wheel's outer radius is negligible, the modal group index was calculated to be 3.62. A narrowband transmission spectrum shown in Fig. 3(b) reveals loaded optical $Q$ factors above 500,000, which is equivalent to a finesse of around 1,112.

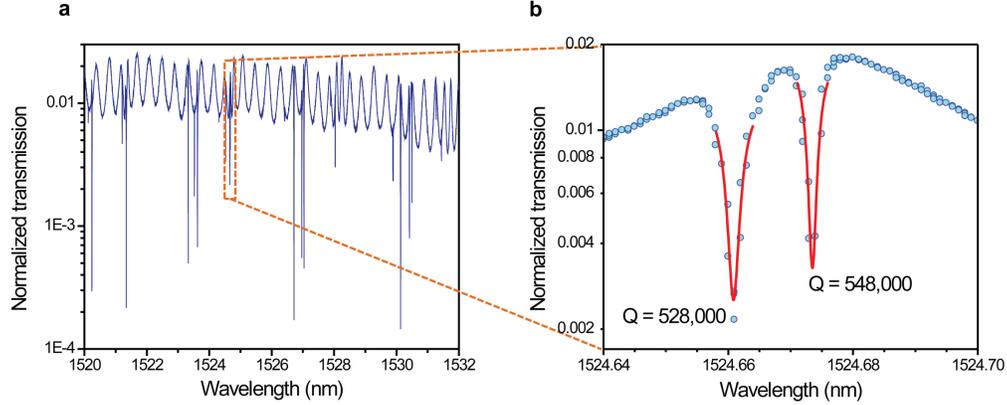

Fig. 3. (a) Normalized optical transmission spectrum of the device. Note that the background fringes result from the two grating couplers, each of which has a coupling loss of around 8 dB. The high-extinction dips correspond to the wheel's optical whispering gallery modes with a free spectral range of 3.39 nm. (b) Zoomed-in transmission spectrum showing the loaded optical $Q$ factors exceeding 500,000.

We measured the wheel's mechanical modes by setting the input laser wavelength at the maximum slope of an optical resonance and recording the noise spectrum of the optical transmission. The wheel resonator's mechanical motion causes phase variation of the optical mode, and as a result induces a resonance shift at the frequencies of the mechanical modes. Therefore the noise spectrum of the optical transmission contains the signature of the wheel's vibrational modes because of the resulting intensity modulation [19]. To compensate the insertion loss introduced by the grating couplers and enhance the signal, the transmitted light was sent through a fiber preamplifier before reaching the GHz photodetector. The detected signal was then sent to an electrical spectrum analyzer to get the radio-frequency (RF) power spectral density.

Figure 4 shows the RF spectrum of the mechanical modes of the wheel resonator measured in air. The data were recorded below the threshold of mechanical self-oscillations [19]. The laser intensity was low enough that optomechanical gain from the wheel resonator is negligible. Since the wrap-around coupling waveguide lies in the same plane as the wheel resonator, the in-plane mechanical motions have the strongest optomechanical coupling. The four dominating groups of the modes shown in Fig. 4(a) are attributed to the designed radial contour modes.

The frequencies of the strongest peak of each order are in very good agreement with their simulated values (see Table 1). In each order of the radial contour modes, the other weak peaks seen around the strongest one are attributed to the higher-order azimuthal modes due to the broken radial symmetry because of the spokes. These modes were also identified in the FEM simulation. Figure 4(b) displays the enlarged spectra for the four radial contour modes, with their fitted mechanical $Q$ factors. The fundamental mode possesses a lower $Q$ factor (around 1,500) than all the higher-order modes (around 4,000). This can be attributed to the fact that the fundamental mode is a radial breathing mode and thus experiences larger viscous damping from the air because of the large collective radial movement of the ring. The higher-order modes, by contrast, are more immune to this type of damping loss because their relative motions exist only inside the ring. Among the higher-order modes, the 3rd mode (921 MHz) exhibits the highest $Q$ factor approaching 4,100. This suggests that our optimal design targeted for this mode indeed results in a lower clamping loss as the spokes are attached to the mode's nodal point.

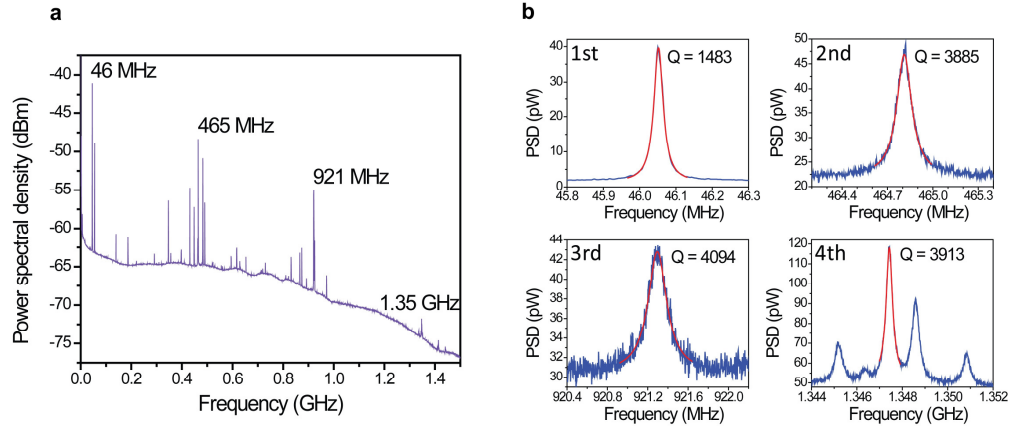

Fig. 4. (a) Full-range RF power spectral density of the device, showing the transduced mechanical modes. The first four orders of radial contour modes at 46 MHz, 465 MHz, 921 MHz, and 1.35 GHz are distinct. (b) Zoomed-in RF spectrum at each radial contour mode showing the mechanical $Q$ factor. PSD: power spectral density.

Our mechanical $Q$ factor of 3,900 for the 1.35 GHz mode is about 30 times higher than that recently reported for GaAs disks at the same frequency level [10]. This significant $Q$ enhancement can be attributed to our nonintrusive suspension design, which minimizes the support losses by the two-stage isolation strategy. The frequency–$Q$ product of $5.3 \times 10^{12}$ is very close to the highest reported value in GHz optomechanical resonators operating at room temperature and atmospheric pressure [9].

## 4. Conclusion

We have demonstrated optomechanical resonators, vibrating in mechanical radial contour modes with frequency covering VHF (46.0 MHz) and UHF (465 MHz, 921 MHz, and 1.35 GHz), on a Si integrated photonics platform. The loaded optical $Q$ factor is above 500,000, corresponding to an optical loss rate of $\kappa/2\pi \approx 390$ MHz. This makes those mechanical modes in the UHF range well seated in the resolved-sideband regime ($\kappa/2\Omega_m < 1$) [20]. The mechanical $Q$ factors of those UHF modes are around 4,000 at room temperature and atmospheric pressure. Obtaining high $Q$ factor in GHz optomechanical resonators is not only beneficial to practical sensing applications, but also crucial in optomechanical backaction cooling of the mechanical modes to their quantum ground state. Efforts in further enhancing the $Q$ factor and exploring even higher frequency modes by improving the device structure and measurement techniques are underway.

## Acknowledgments

This work was supported by DARPA/MTO and National Science Foundation CAREER award. W.H.P.P. thanks the Alexander-von-Humboldt Foundation for providing a postdoctoral fellowship. H.X.T. acknowledges support from a Packard Fellowship in Science and Engineering. The authors thank Michael Power and Dr. Michael Rooks for assistance in device fabrication.